\documentstyle[11pt,leqno,twocolumn]{article}
 \newcommand{\beq}{\begin{equation}}                       
 \newcommand{\eeq}{\end{equation}}                         
 \newcounter{nt}[section]                                  
 \newcounter{nl}[section]                                  
 \date{}                                                   

\begin{document}


 \title{ \bf On hereditary models of polymers}

\vspace{3mm}
\author
{\sc Monica De Angelis \\
\\
Universit\`{a} Studi di Napoli Federico II, Facolt\`{a} di
             Ingegneria, \\ Dip. Mat. e Appl., via Claudio 21, 80125,
             Napoli; Italy. \\ Tel. 081 7683387 e-mail modeange@unina.it}

 \maketitle

\vspace{3mm}
 \begin{abstract}

\vspace{3mm}
 An equivalence between an integro-differential operator
${\cal M}$  and an
evolution  operator
${\cal L}_n$
 is determined. Owing to this equivalence,  the fundamental solution of ${\cal L} _n$
 is estimated in terms of the fundamental solution
 related to  the third order operator ${\cal L}_1$ whose behaviour is
 now acquired.
Moreover, properties typical of wave hierarchies can be applied to
polymeric materials.
 As an example  the case $n=2$  is considered and results are applied to  {\em the Rouse model} and
 {\em the reptation model } which describe different aspects of polymer
chains.

 \end{abstract}

  \vspace{3mm}

 \section{\hspace*{-6mm}{\Large \bf .\hspace{2mm}  Introduction
}}
 \setcounter{equation}{0}

 \hspace{3mm}

The creep and relaxation processes related to the viscoelastic
behavior of many actual polymeric materials are specified by means of
memory functions like:

\beq          \label{11}
g_n(t)= \sum_{h=1}^{n} B_h \ e^{- \beta_h t },
\eeq

\vspace{3mm}
\noindent
where $n$,   $ B_h$ and $\beta_h$  depend on the polymer physics and
 are  determined in order to fit the
experimental curves for $g_n(t)$ with a required approximation
[\ref{f}-\ref{de}].

Moreover, according to the well-known Muntz and Schwartz'z theorem
[\ref{b}-\ref{s}], every continuous spectrum can be uniformly
represented by Dirichlet polynomials.

\noindent
In this paper, an integro - differential operator  ${\cal M}$ related to
(\ref{11})  is considered and a {\em conditioned} equivalence
between  ${\cal M}$  and the  $(2+n)$ order operator:

\beq                  \label{12}
 {\cal L}_ n =   \partial_{t}^{(n)} (
\partial_{tt}-c_n^2 \partial_{xx} ) \ +
 \eeq
\hspace*{2cm}\[  \ a_{n-1}  \partial_{t}^{(n-1)}(
\partial_{tt}-c_{n-1}^2 \partial_ {xx} ) +.. a_{0} \ (
\partial_{tt}-c_{0}^2 \partial_ {xx} )  \]

\vspace{3mm}
\noindent
where  $ a_k \ (k=0..n-1)$ are positive constants,
 is extablished.

\vspace{2mm}
By  this equivalence whatever $n$ may be the fundamental solution of
(\ref{12})
is explictly  determined and an estimate in terms of the fundamental
solution related to the third order operator is achieved, too.

 As applications, when  $n=2$, results are applied to polymer chains
 and  {\em the Rouse model} and
 {\em the reptation model } are considered.

  \vspace{3mm}

 \section{\hspace*{-6mm}{\Large \bf .\hspace{2mm}  Statement of the
 problem}}
 \setcounter{equation}{0}

 \hspace{3mm}

Let ${\cal B} $  a linear, isotropic,
 homogeneous system and let  ${\underline u}(x,t) \underline{i}$  the displacement
field from an underformed reference configuration
  ${\cal B}_0$.

Indicating by  $\sigma $ and
$\varepsilon $ the only non vanishing components of the stress and
the strain tensors,
 the
 constitutive relation is

\vspace{3mm}
 \beq                                   \label{21}
\sum _{k=0}^n a_k \ \partial_t^{k} \sigma =\sum _{k=0}^n \alpha_k
\ \partial_t^{k} \varepsilon
 \eeq

\vspace{3mm}
\noindent
with $a_k, \alpha_k$ constant $(a_n,\alpha_n \neq 0)$.

\noindent
 So, if    $\underline{\bf{f}}= f  \underline{\bf {i}}$  is the known body force, $\rho_0 $
 denotes the mass density in ${\cal B}_0$, and      $c_k= \alpha _k /
 \rho_0 a_k, $
 the one dimensional
linear motions of
 ${\cal B}$ are described by higher order  equation [\ref{c}]:

\beq          \label{22}
{\cal L }_n u = F
\eeq

\vspace{2mm}
\noindent
 where

\beq          \label{23}
{\cal L }_n  = \sum_{k=0}^{n} a_k  \partial _t^{(k)}  (\partial_{tt}-c_k
\partial_{xx})
\eeq

\noindent
and
\noindent

\beq                \label{24}
 F=(1 / \rho_0 ) \ \sum _{k=0}^n a_k
\ \partial_t^{k} f.
\eeq

\vspace{2mm}
In (\ref{23}) constants  $c_k$ are the characterized speeds depending on the
materials properties of the medium and in many physical problems it
results
  $c_0^2 < c_{1}^2..< c_{n-1}^2 < c_n^2$     and so the
  equation (\ref{12}) is typical of {\it
 the wave hierarchies}. [\ref{w}].

When $n=1$,  (\ref{23}) turns into a strictly hyperbolic third -
order operator which models the evolution of the Standard Linear Solid
 (S.L.S.) [\ref{hu}-\ref{h}]
 and  its behaviour has been already discussed in [\ref{r3}]. The
fundamental solution ${\cal E}_1$ has been explicitly determined together with
maximum theorems and boundary layer estimates.

Moreover, if   $J(t)$ denotes the creep-compliance, the behavior of most
viscoelastic media is fairly well modelled by linear hereditary
equations like:

 \beq                                   \label{25}
\varepsilon (t) \ = \ J(0) \ \ \ [\sigma(t) +
\eeq
\hspace* {2cm} \[ + \int_{-\infty}^t \dot J (t-\tau)
\sigma (\tau) d\tau]. \]

\vspace{2mm}
According to {\it fading memory} hypotheses [\ref{g}-\ref{g2}], $\dot J(t)$ is a
positive fast decreasing function and in many real
materials as polymers rubbers or bitumines, it
 is representated by means of chains of S.L.S. elements in series or
parallel [\ref{d}-\ref{de}]. In the series case,
 one has:

 \beq                                   \label{26}
 J_n(t) = \ J_n(0) \ \ [  \ 1  +
\eeq
\hspace* {2cm} \[  \  + \ \sum _{k=1}^n \ \ \frac{B_k}{\beta_k}(1-
 e^{-\beta_kt})], \]

\vspace{2mm}
\noindent
where $n$ is the number of elements in the chain,
 $J_n(0)  $ denotes the
elastic compliances and  $\tau _k= \beta_k^{-1}$
are the characteristic times .

Let consider equations of one dimensional motions of  ${\cal B}$:

\beq        \label {27}
\rho_0 \; u_{tt} = \sigma  + f , \ \ \   \varepsilon= u,
\eeq

\vspace{3mm}
\noindent
by means of (\ref{25}) - (\ref{27})  the following integral constitutive
 equation is deduced:

\vspace{3mm}
 \beq                                   \label{28}
 {\cal M}u  = c^2 u_{xx}- u_{tt} -
\eeq
\hspace* {2cm} \[ - \int_0^t g(t-\tau)u_{\tau\tau}
 d\tau = -F_*(x,t) \]

\noindent
where

 \beq                                   \label{29}
c^2 = [\rho_0 J_n(0)]^{-1} , \ \
\eeq
\hspace* {2cm} \[F_* =c^2 [J_n(0) f
+\int_{-\infty}^0 \dot J_n (t-\tau)
\sigma_x (\tau) d\tau]. \]

\noindent
and

 \beq                                   \label{210}
g \ = \ g_n(t)= \sum _{k=1}^n B_k e ^{-\beta _k t} \ = \dot J_n(t)
/J_n(0).
 \eeq

\vspace{2mm}

\vspace{3mm}\noindent
In this memory function, $n $ is quite
 arbitrary and constants $B_k $  and frequencies $\beta_k$
are such that:

\beq            \label{211}
0 < \beta _1 < \beta_2
...< \beta_n;
\eeq
\hspace* {2cm} \[  B_k > 0  \ \ \forall k =1,2....n. \]

\vspace{3mm}
Whatever n may be, the fundamental solution $E_n$ of operator ${\cal M}$
 has been explictly determined [\ref{r1}-\ref{r2}]. Moreover, let
 $E_1$  the fundamental solution related  to an
 appropriate S.L.S.  ${\cal B}_1^* $ defined by:

\beq             \label{212}
g_1 = b \  e^{-\beta_1 t} \ \   \
\eeq

\noindent
with

\beq             \label{213}
\ \ b = \beta_1  \sum_1^n
\frac{B_k}{\beta_k},
\eeq

\vspace{2mm}
\noindent
the following theorem assures that the fundamental solution  $E_n$ can
be rigorously estimated by means of
 $E_1$.

In fact, if
 $\Gamma $ is the open forward
characteristic cone $\{ (t,x) : t >0 \ |x| < ct \}$,
and  $\chi_n =
 \prod_{k=2}^n \ (\frac{B_k}{\beta_1})^2$,
then the following theorem holds:

\vspace{4mm}
{\bf Theorem 1.1} - {\em If the memory function is given by }
(\ref{210})
(\ref{211}), {\em
then the fundamental solution} $E_n$ {\em of} $ {\cal M}$  {\em is a never negative }
$C^{\infty}(\Gamma)$ {\em function and
 it
satisfies the estimate}:

\beq                     \label {214}
 0< E_n(\beta_1..\beta_n,B_1..B_n)<
\eeq
\hspace* {2cm} \[< \chi_n \ \
 E_1(\beta_1, b
),\]

\vspace{2mm}
\noindent
{\em everywhere in the cone  } $\Gamma$
{\em and
whatever } $n$ {\em may be}. \hbox {} \hfill \rule {1.85mm} {2.82mm}

\vspace{3mm}
 \section{  \hspace*{-6mm}{\large \bf .\hspace{2mm}  Conditioned equivalence  between operators ${\cal L}_n $ and
 ${\cal M}$}}
 \setcounter{equation}{0}

 \hspace{3mm}

Let be null the initial data related to (\ref{22}) and (\ref{28}) and
let

\vspace{2mm}
\beq                                   \label{31}
\mu _k = a_k/a_n,
\eeq
\hspace* {2cm} \[\lambda_k = a_kc_k/a_nc_n
\ \ \ (k=0,..n).\]

\vspace{2mm}\noindent
 Appling the Laplace transform and  the
 polinomial identity one has $c_n=c^2 $  and:

\vspace{2mm}
\beq                    \label{32}
\hspace*{5mm} \left \{
\begin {array}{l}
 \lambda_0 =
 \beta_1 \beta_2..\beta_n \\
 ..............................\\
 \lambda_{n-2}=
 \beta_1 \beta_2+ \beta_1 \beta_3+..\beta_{n-1} \beta_n  \\
\\
 \lambda_{n-1}=    \beta_1+....+\beta_n
\end {array}
\right.
\eeq

\vspace{3mm}
\noindent
So, owing to (\ref{211}), all the $\lambda_k$'s are positive. Further, as
for $\mu_k$, one has:

\beq                    \label{33}
\hspace*{5mm} \left \{
\begin {array}{l}
\mu_0= \lambda_0 +   B_1
(\beta_2..\beta_n)+... \\
\hspace*{1cm} ...+B_n(\beta_1..\beta_{n-1})\\
..............................\\
\mu_{n-2} = \lambda_{n-2} +  B_1
(\beta_2+..+\beta_n) \\

\hspace*{1cm} ...+B_n(\beta_1+..+\beta_{n-1})  \\
\\
\mu_{n-1} = \lambda_{n-1} +    B_1+....+B_n
\end {array}
\right.
\eeq

\vspace{2mm}\noindent
 and (\ref{211}), (\ref{33}) imply too:

\beq     \label{34}
0<\lambda_k<\mu_k \ \ (k=0,..n-1).
\eeq

\vspace{2mm}\noindent
So, for $ k=0,...n-1$, one has:

\beq                        \label{35}
 0< c_k < c_n = c^2.
\eeq

\vspace{2mm}\noindent
At last, by (\ref{32}),(\ref{33}), it follows:

\beq                 \label{36}
\frac{\lambda_0}{\mu_0} \, < \,\frac{\lambda_1}{\mu_1} \,<
 \, \frac{\lambda_{n-1}}{\mu{n-1}}
\eeq

\vspace{2mm}\noindent
and hence:

\beq                                   \label{37}
0 <  c_0 < c_1 ....\ < c_n.
\eeq

\noindent
As consequence, the following property holds:

\vspace{4mm}
{\bf Property 3.1} {\em Hypotheses of fading memory} (\ref{210})
(\ref{211}) {\em imply
that the differential operator} (\ref{23}) {\em is typical of wave
hierarchies}.   \hbox {} \hfill \rule {1.85mm} {2.82mm}

\vspace{4mm}
Vice versa
 when the differential equation  (\ref{22}) is prefixed, to
obtain the dual hereditary equation  (\ref{28}) with a memory function
$ g_n(t)$ satisfying (\ref{210}), (\ref{211}), appropriate restrictiones on
the constants $a_k, c_k $ must be imposed.

\vspace{3mm}
{\bf Example 3.1}

\vspace{2mm}\noindent
When $n=2$, one has  $c^2=c_2, B_0=1$. Further
$\beta_ 1 , \beta_2 $ are real iff:

\beq                          \label{38}
 \hspace*{3mm}\omega^2 = (a_1c_1)^2 - 4 (a_0c_0)(a_2c_2)>0.
\eeq

\noindent
So, beeing:

\vspace{2mm}
\beq                          \label{39}
\beta_1 = \frac{1}{2a_2c_2} (a_1c_1 - \omega) ,
\eeq
\hspace* {2cm} \[  \beta_2 =\frac{1}{2a_2c_2} (a_1c_1 + \omega), \]

\vspace{2mm}\noindent
it results $0<\beta_1<\beta_2$ .

\noindent
Moreover as for $B_i (i=1,2),$ one has:

\vspace{2mm}\beq                          \label{310}
B_i = \frac{(-1)^{i-1}}{\omega} [a_0 (c_2 - c_0 ) -
\eeq
\hspace* {2cm} \[ - a_1 \beta_i
(c_2-c_1 )],  \ \ \ \\ (i=1,2), \]

\vspace{2mm}\noindent
and hence, $ B_1>0, B_2>0 $ iff

\vspace{2mm}
\beq                          \label{311}
\beta_1 < \frac{a_o}{a_1} \ \ \frac{c_2-c_0}{c_2-c_1} < \beta_2.
\eeq

\noindent
Therefore, the fourth-order operator:

\beq                          \label{312}
a_2 (u_{tt}-c_{2} u_ {xx} )_{tt} \ \ +
\eeq
\hspace* {2cm} \[ + a_1 ( u_{tt}-c_{1} u_{xx})_{t}+a_0
(u_{tt}-c_0 u_{xx}) \]

\vspace{2mm}\noindent
can be analyzed by integral operator ${\cal M} $ (\ref{28}) with
conditions (\ref{210}),
 (\ref{211}) and when   the
constants  $a_k, \ c_k$  satisfy  (\ref{38})  and  (\ref{311}).
\hbox {} \hfill \rule {1.85mm} {2.82mm}

\vspace{3mm}
 \section{ \hspace*{-6mm}{\Large \bf .\hspace{2mm}  Polymeric
 materials}}
 \setcounter{equation}{0}

 \hspace{3mm}

Polymeric materials are very flexible like rubber and are easily
formed into fibres, thin films, etc.
 Moreover, the liquid state composed only of polymers (polymer melt) is
an important state for industrial uses where polymeric materials are
processed into various plastic products such as gaskets, seals,
flexible joints, vehicle tires, etc..

\noindent
 Also the durability is a requirement imposed on polymers and  polymeric
composites and
 the interest for  future development of  these materials is increasing more and more.

\noindent
A large literature treats with polymer physics and as for
 viscoelastic theories, two models which describe different aspects of
 the polymer chains, have met with reasonable success:
 {\em the Rouse model} and {\em the reptation
model} [\ref{d}-\ref{de}].

In both cases the memory function $g_n(t) $ assumes a form like
(\ref{11}).

In fact in {\em the Rouse model}  function $g_n(t)$ is given by:

\beq             \label{41}
g_n(t) = k_1 \sum _{h=1}^n  e^{2h^2 \frac{t}{\tau_1}}
\eeq

\noindent
where
 the relaxation time $\tau_1$  can be calculated by means of
 experimental
results.[\ref{de}].

When the  viscoelastic behaviour is represented by
 {\em the reptation model},
 as times increases, the stress function decreases with a relaxation
time $\tau_d$, and one has:

\vspace{3mm}
\beq             \label{42}
g_n(t) = k \sum     \frac{1}{h^2} \ \ e^{-h^2
\frac{t}{\tau_d}},
\eeq

\vspace{3mm}
\noindent
where $h $ ranges over odd integer, the constant  $k$ depends
on the polymer physics and the value of the {\em reptation} time  $\tau_d$ can be fixed according
to
 elasticity experiments [\ref{f}].

\noindent
So, if one considers the first two steps in  {\em
the reptation model}, it results:  $B_1=k, \ \ B_2 =B_1/9, \ \ \beta_1=1/\tau_d, \ \
\beta_2= 9\beta_1.$  Consequently the operator (\ref{312}) is
characterized by constants:

\beq               \label{43}
\hspace*{2mm}
 \left\{
\begin {array}{ll}
c_0= c^2 \
\frac{81}{81+82k \tau_d} & a_0=1+ \frac{82}{81} \, k \tau_d \\

\\
 c_1= c^2 \ \frac{9}{9+k\tau_d} & a_1=\frac{10
 \tau_d^2}{9}(\frac{1}{\tau_d}+\frac{k}{9}) \\

\\
 c_2= c^2  & a_2 =\frac{\tau_d^2}{9}\\
\end {array}
\right.
\eeq

\vspace{4mm}
\noindent
Analogously, in {\em the Rouse model}, beeing $B_1=B_2=k_1,
\ \ \beta_1=2/\tau_1, \ \  \beta_2=4\beta_1$, one has:

\vspace{3mm}
\beq               \label{44}
\hspace*{4mm} \left\{
\begin {array}{ll}
 c_0= c^2
\frac{8}{8+5 k_1 \tau_1}  & a_0 = 1+ \frac{5 k_1 }{8}\tau_1  \\
\\
c_1= c^2  \frac{5}{5+ k_1 \tau_1} & a_1 = \frac{\tau_1^2}{16} (2 k_1
+\frac{10}{\tau_1})\\
\\
 c_2= c^2 &a_2=\frac{\tau_1^2}{16}
\end {array}
\right.
\eeq

\vspace{4mm}
The {\em wave hierarchies} defined by
 (\ref{43}) or (\ref{44})  are governed by the operator   ${\cal L}^*_1    $
of the Standard Linear Solid defined, respectively, by:

\beq               \label{45}
\hspace*{4mm} \left\{
\begin {array}{ll}
c_0= c^2 \
\frac{81}{81+82k \tau_d} &
 c_1= c^2  \\
 \\
a_0=1+ \frac{82}{81} \, k \tau_d &
\eta =\frac{81 \tau_d}{81+82 k \tau_d}, \\
\end {array}
\right.
\eeq

\beq               \label{46}
\hspace*{2mm} \left\{
\begin {array}{ll}
c_0= c^2
\frac{8}{8+5 k_1 \tau_1}  &
 c_1= c^2 \\
 \\
 a_0 = 1+ \frac{5}{8} k_1 \tau_1 &
\eta = \frac{4 \tau_1}{8+5 k_1 \tau_1}.
\end {array}
\right.
\eeq

\vspace{3mm}
{\bf Remark 3.1-} As swowed, memory function $g_n(t)$ can depend on
$h^2$. So the approximation to the two first terms appears to be
reasonable. However, in many article the model is limited to a single
relaxtion time (see. f.i.[\ref{im}]).
 \hbox {} \hfill \rule {1.85mm} {2.82mm}

\vspace{3mm}
\begin{center} {\Large \bf {References}}
\end{center}

\vspace{3mm}
\begin{enumerate}

\item Ferry,   J. D.: Viscoelastic properties of polymers.
New York. Wiles, 1961 \label{f}

\item Doi, M. See, H.: Introduction to Polymer Physics.
 Oxford:  Clarendon
Press, 1977 \label {d}

\item Doi, M.  Edwards S. F.:  The theory of Polymer
Dynamics. Oxford: Clarendon Press, 1986 \label {de}

\item Sunyer Balaguer, F. : Approximation of functions by
linear combinations of exponentials.
Collect. Math. {\bf17}, 145-177, (1965). \label{b}

\item  Schwartz, L.: Etude des sommes d'exponentielles. Act.
Scient. et Industr. {\bf 959}, Hermann, Paris (1959). \label{s}

\item  Christensen, R. M.: Theory of Viscolasticity. New York : Academic
Press, 1971 \label{c}

\item  Whitham, G. B.: Linear and Non linear Waves. New York : John Wiley e
Sons, 1974 \label{w}

\item Hunter, S.C.:  Mechanics of Continuous media.
England: Ellis Horwood, 1976 \label{hu}

\item  Haupt, P.: Continuous Mechanics and theory of
Materials, Springer 2000 \label{h}

 \item  Renno, P.:  On a Wave Theory for the Operator
$\varepsilon \partial_t(\partial_t^2-c_1^2
\Delta_n)+\partial_t^2-c_0^2\Delta_n$. Ann. Mat. pura e Appl.,{\bf
136(4)}
355-389 (1984). \label{r3}

\item  Graffi. D. : On the fading memory. Applicable
 Analysis, {\bf 15} (1983) 295-311 \label{g}

  \item Graffi. D.: Mathematical models and waves in linear
viscoestacity. Euromech Colloquium 127 on {\it Waves propagation in
viscoelastic media}, Pitman Adv. Publ. Comp. Research notes in
Math.,{\bf 52}, 1-27 1980 \label {g2}

 \item Renno, P.: On the Cauchy problem il linear
 viscoelasticity. Ren. Acc. Naz. Lincei,  VIII vol. {\bf LXXXV} (1983)
 \label{r1}

 \item Renno, P.: On some viscoelastic models.  Ren. Acc. Naz.
 Lincei, VIII vol. {\bf LXXV} (1983) \label {r2}

\item Ianniruberto,G.,  Marrucci G.: A simple constiyutive
equation for entangled polymers with chain stretch. J. Rheol.{\bf45}(6)
1305- 1318 (2001) \label{im}

\end{enumerate}

\end{document}